\documentclass[preprint,12pt]{elsarticle}
\usepackage{graphicx}
\usepackage{colordvi}
\journal{Optics Communications}
\begin{document}
\begin{frontmatter}

\title{Effect of quintic nonlinearity on modulation instability in coupled nonlinear Schr\"odinger systems}
\author[a]{B. B. Baizakov \corref{*}}
\ead{baizakov@uzsci.net}
\cortext[*]{Corresponding author.}
\author[b]{A. Bouketir}
\author[c]{S. M. Al-Marzoug}
\author[c]{H. Bahlouli}
\address[a]{Physical-Technical Institute, Uzbek Academy of Sciences, 100084, Tashkent, Uzbekistan}
\address[b]{Dammam Community College, King Fahd University of Petroleum and
Minerals, Dhahran, 31261, Saudi Arabia}
\address[c]{Physics Department, King Fahd University of Petroleum and
Minerals, and Saudi Center for Theoretical Physics, Dhahran, 31261, Saudi Arabia.}

\begin{abstract}
Modulation instability (MI) in continuous media described by a
system of two cubic-quintic nonlinear Schr\"odinger equations (NLSE)
has been investigated with a focus on revealing the contribution of
the quintic nonlinearity to the development of MI in its linear and
nonlinear stages. For the linear stage we derive analytic expression
for the MI gain spectrum and compare its predictions with numerical
simulations of the governing coupled NLSE. It is found that the
quintic nonlinearity significantly enhances the growth rate of MI
and alters the features of this well known phenomenon by suppressing
its time-periodic character. For the nonlinear stage by employing a
localized perturbation to the constant background we find that the
quintic nonlinearity notably changes the behavior of MI in the
central oscillatory region of the integration domain. In numerical
experiments we observe emergence of multiple moving coupled solitons
if the parameters are in the domain of MI. Possible applications of
the obtained results to mixtures of Bose-Einstein condensates and
bimodal light propagation in waveguide arrays are discussed.
\end{abstract}


\end{frontmatter}

\section{Introduction}

Modulation instability (MI) represents one of the commonly observed
phenomena in nonlinear physics
\cite{zakharov,abdullaev,biondini2018}. It shows up as an
exponential growth of small amplitude perturbations of the plane
wave solutions of the governing nonlinear equation and is
responsible for a variety of pattern formation phenomena, including
formation of soliton trains. While substantial research on MI has
been accomplished since the first studies in nonlinear optics
\cite{ostrovskii}, plasma physics \cite{taniuti}, hydrodynamics
\cite{benjamin}, electrical transmission lines~\cite{remoussenet},
and more recently in Bose-Einstein condensates (BEC) \cite{mi_bec},
the field still appears to be rich for interesting new discoveries
\cite{zakharov2013,biondini2016}. A prominent example is the so
called rogue waves in continuous \cite{akhmediev} and discrete
\cite{tsironis} nonlinear systems, which develop due to the MI. A
more recent example is reported in \cite{ferrier-barbut2018}, where
the MI was a precursor to generation of multiple quantum droplets in
dysprosium dipolar BEC. Relevance to extreme ocean waves, generation
of high intensity electromagnetic pulses in optical media and new
discoveries in physics has been the motivation for sustained
interest in MI over the years.

MI was employed for generation of bright soliton trains in optical
fibers \cite{hasegawa} and BEC~\cite{strecker,nguen2017}. Inclusion
of higher order nonlinearities gives raise to novel phenomena, not
observed in media with only cubic nonlinearity
\cite{fabrelli2017,saha2012}. The role of cubic-quintic nonlinearity
in phase separation of a two-component BEC loaded in a deep optical
lattice was reported \cite{baizakov2009}. The essential difference
between MI in discrete and continuous systems is that, in the former
case emerging solitons are pinned by the lattice due to a
Peierls-Nabarro potential \cite{kivshar1993}, while in the latter
case solitons can move and interact with each other.

Our objective in this work is to study the MI in continuous media
with cubic-quintic nonlinearity, described by a system of two
coupled nonlinear Schr\"odinger equations (NLSE). We mainly focus on
revealing the contribution of quintic nonlinearity on the MI gain
spectrum. To get insight into nonlinear stage of MI in this system,
we employ new evidences for the universal behavior of waves in MI
supporting media, reported in Ref. \cite{biondini2018,biondini2016}.
Theoretical predictions of these works for the nonlinear stage of MI
initiated by localized perturbation has been confirmed in a recent
fiber optic experiment \cite{kraych2018}.

The paper is organized as follows. In the Sec. 2 we introduce the
coupled system of NLSE and perform a linear stability analysis of
their plane wave solutions. Sec. 3 is devoted to numerical
simulations of the development of MI. Coupled solitons emerging from
MI of flat-top localized states is considered in Sec. 4. In the last
Sec. 5 we summarize our findings.

\section{Model and linear stability analysis}

The mathematical model is based on the coupled system of two NLSE
with cubic-quintic nonlinearity, introduced in Ref. \cite{maimistov}
\begin{eqnarray}\label{gpe}
i\partial_t\psi_{j} &+& c_j\partial_{xx}\psi_{j} + \lambda(|\psi_{j}|^2+\beta|\psi_{3-j}|^2)\psi_{j}+ \nonumber \\
&+&
\gamma(|\psi_{j}|^4+2\alpha|\psi_{j}|^2|\psi_{3-j}|^2+\alpha|\psi_{3-j}|^4)\psi_{j}
= 0, \qquad j=1,2.
\end{eqnarray}
A relevant physical system in nonlinear optics can be represented by
the above equations to describe the propagation of two orthogonal
polarizations of light in materials with third- and fifth-order
susceptibilities. In nonlinear optics $\psi_j$ stands for the
amplitude (slow envelope) of the light field, the evolution variable
$t$ (usually denoted by $z$), has the meaning of propagation
distance, while the variable $x$ (denoted by $t$), stands for time.
As examples of such materials chalcogenide glasses~\cite{smectala}
and nonlinear polymeric materials \cite{lawrence} can be mentioned.
A similar system of equations was also considered in Ref.
\cite{elyutin2009} to describe polarized optical pulses in a medium
with third- and fifth-order nonlinearities.

To be specific, below we consider a two-component BEC confined to a
quasi-1D trap, described by appropriately normalized mean field wave
functions $\psi_1(x,t)$ and $\psi_2(x,t)$. The two coefficients
$c_j$ account for possible different masses of atomic species in the
condensate mixture. We use the auxiliary coefficients $\lambda$ and
$\gamma$ to switch between the purely cubic ($\lambda = \pm 1$,
$\gamma = 0$), purely quintic ($\lambda = 0$, $\gamma = \pm 1$), and
mixed ($\lambda = \pm 1$, $\gamma = \pm 1$) nonlinear interactions
between the two components. The strength and sign of the cubic and
quintic inter-component interactions are changed via the
coefficients $\beta$ and $\alpha$, respectively.

Although the majority of research on BEC is performed in the
framework of Eqs. (\ref{gpe}) with cubic nonlinearity, in some cases
the contribution of higher order nonlinearities become essential. In
particular, three-body effects in BEC, responsible for the quintic
nonlinearity, become important when the density of the gas is high.

Below we investigate the linear stability of nonlinear plane wave
solutions of following form
\begin{equation}\label{sol}
\psi_j(x,t)=a_j\exp[i(k_j x - \omega_j t)], \quad j = 1,2,
\end{equation}
where $a_j, k_j, \omega_j$ are the amplitudes, wave numbers and
frequencies of the two plane waves, respectively.

Substitution of Eqs. (\ref{sol}) in  Eqs. (\ref{gpe}) gives the
following dispersion relations
\begin{equation}\label{disp}
\omega_j= c_j k_j^2 - \lambda(a_j^2 + \beta a_{3-j}^2) - \gamma
(a_j^4 + 2\alpha a_j^2 a_{3-j}^2 + \alpha a_{3-j}^4), \quad j =
1,2.
\end{equation}
Then we impose a weak modulation on the plane waves
\begin{equation}\label{psi}
\psi_j(t, x)=(a_j + \xi_j) \exp[i(k_j x - \omega_j t)], \quad j =
1,2.
\end{equation}
The perturbations $\xi_j$ will have the form
\begin{equation}\label{pert}
\xi_j = u_j \exp[i(q x - \Omega t)] + v_j \exp[-i(qx - \Omega t)],
\quad j = 1,2,
\end{equation}
where $u_j, v_j$ are the amplitudes of the weak modulation, $q$ is
the modulation wave number and $\Omega$ is its frequency. By
inserting Eqs. (\ref{pert}) into Eqs. (\ref{gpe}) and performing
linearization we end up with an eigenvalue problem for the
perturbation, whose nontrivial solution is associated with the
following condition
\begin{equation}\label{eigen}
\left| \begin{array}{cccc}
L_{1,p} + \Omega & \Lambda_1 & \Lambda_{12} & \Lambda_{12}  \\
\Lambda_1 & L_{1,m} - \Omega & \Lambda_{12} & \Lambda_{12} \\
\Lambda_{21} & \Lambda_{21} & L_{2,p} + \Omega & \Lambda_{2} \\
\Lambda_{21} & \Lambda_{21} & \Lambda_{2} & L_{2,m} -\Omega
\end{array} \right|=0,
\end{equation}
where
\begin{eqnarray}
\Lambda_{j,3-j} &=& a_{j}a_{3-j} \left[\beta \lambda + 2\alpha\gamma(a_{j}^2 +a_{3-j}^2)\right], \label{elements1} \\
\Lambda_j &=&a_{j}^2\left[\lambda + 2\gamma(a_{j}^2 + a_{3-j}^2\alpha)\right], \label{elements2}\\
L_{j,p}&=& \Lambda_j - c_j \, q(2k_j+q), \label{elements3} \\
L_{j,m}&=& \Lambda_j + c_j \, q(2k_j-q), \label{elements4} \quad
j=1,2.
\end{eqnarray}
The eigenvalue problem (\ref{eigen}) yields the following equation
for the modulation frequency~$\Omega$
\begin{equation}\label{omegaequa}
\Omega^4 + p_3\Omega^3 + p_2\Omega^2 + p_1\Omega + p_0=0,
\end{equation}
where $p_i$ are coefficients depending on the parameters $a_j$,
$k_j$, $c_j$, $q$, $\alpha$, $\beta$, $\gamma$ and $\lambda$. Below
we consider the simplest case of uniform equal amplitude plane waves
$a_1=a_2=a$ with zero wave numbers $k_1=k_2=0$. Then according to
Eqs. (\ref{elements3})-(\ref{elements4}) $L_{1,p}=L_{1,m}=L_1$,
$L_{2,p}=L_{2,m}=L_2$, which leads to $p_1=p_3=0$. The expressions
for two remaining coefficients are also simplified
\begin{eqnarray}
p_0 &=& (L_1^2-\Lambda_1^2)(L_2^2-\Lambda_2^2)-4\Lambda_{12}\Lambda_{21} (L_1-\Lambda_1)(L_2-\Lambda_2), \label{p0} \\
p_2 &=& \Lambda_1^2+\Lambda_2^2-L_1^2-L_2^2. \label{p2}
\end{eqnarray}
Then the characteristic Eq. ({\ref{omegaequa}}) is reduced to the
following form
\begin{equation}\label{gaineq}
\Omega^4 + p_2\Omega^2 + p_0=0. \\
\end{equation}
The growth rate of modulation instability (G) as a function of
modulation wave number (q) can be straightforwardly calculated
from the last equation
\begin{equation}\label{gainst}
G = |{\rm Im} \Omega|=|{\rm Im} \sqrt{-\frac{1}{2}(p_2 +
\sqrt{p_2^2-4 p_0})}|,
\end{equation}
where the dependence of the quantities $p_0$ and $p_2$ on the wave
number (q) and other parameters of the system is given by Eqs.
(\ref{elements1})-(\ref{elements4}) and Eqs. (\ref{p0})-(\ref{p2}).

\begin{figure}[htb]
\centerline{\includegraphics[width=8cm,height=6cm,clip]{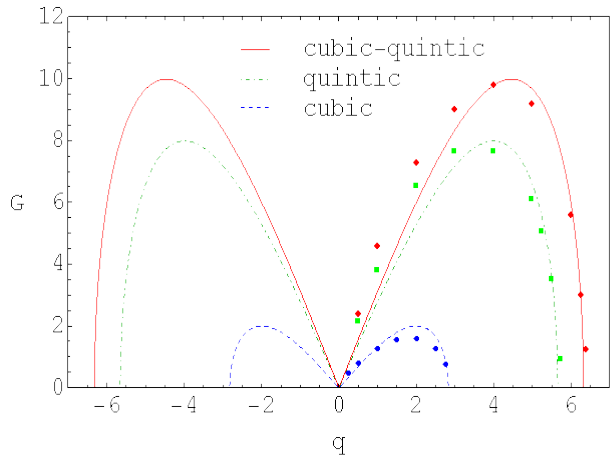}}
\caption{The growth rate of MI as a function of modulation wave
number. When only the cubic nonlinearity is in action ($\lambda=1,
\beta=1, \gamma=0$), the modulation of the plane waves with $|q| <
2.7$ leads to its instability (blue dashed line). The quintic
nonlinearity ($\lambda=0,\gamma=1,\alpha=1$) leads to instability in
a wider region of modulation wave numbers $|q| < 5.6$ (green
dot-dashed line). The combined action of both nonlinearities
($\lambda=1, \beta=1, \gamma=1,\alpha=1$) further expands the domain
of instability $|q|<6.3$ (red solid line). The wave number,
corresponding to the maximum growth rate, also shifts to greater
values. Numerical values of MI growth rates, obtained from Eq.
(\ref{gpe}) with weakly modulated ($u_j = v_j = 0.01$) plane wave
initial conditions and using Eq. (\ref{gain}), are represented by
symbols. For negative values of the modulation wave number $q$ the
data are symmetric with respect to the origin, hence not shown.
Parameter values: $c_1=c_2=0.5,\, a_1=1.0, \, a_2=0.999$. }
\label{fig1}
\end{figure}

Figure \ref{fig1} illustrates the contributions of the cubic and
quintic nonlinearities, as well as their combined action, on the
growth rate of MI, according to Eq. (\ref{gainst}). As expected, the
higher order nonlinearity expands the domain of instability and
leads to faster growth of perturbations. In the same figure we show
through different symbols the gain factor $G_{num}$, obtained from
numerical simulations of the governing Eq. (\ref{gpe})
\begin{equation}\label{gain}
G_{num} = \frac{{\rm ln}(A_{out}/A_{in})}{t_s},
\end{equation}
where $A_{in}$, $A_{out}$ are the amplitudes of perturbation at the
initial ($t=0$) and final ($t=t_s$) stages of the simulation. Using
amplitude perturbations with different spatial frequencies $q$ we
get different growth rates, thus obtaining the gain spectrum shown
in Fig. \ref{fig1}.

\section{Numerical results}

Numerical simulations are performed by standard split-step fast
Fourier transform method \cite{agrawal} using 2048 Fourier modes
\cite{press} within the integration domain of length $x \in [-20
\pi, 20 \pi]$, and the time step was $dt=0.001$. The spacial
frequency of weak modulation of the plane waves $q$ has been
selected so that the periodic boundary conditions for Eqs.
(\ref{gpe}) is satisfied (i.e. the integer number of modulation wave
periods fits the integration domain). When considering the flat-top
soliton initial conditions, we put absorbers at the domain
boundaries \cite{berg} to prevent the interference of linear waves
(emitted by solitons in the central part and subsequently reflected
from the domain boundaries) with emerging localized structures. In
fact, the absorbers emulate the unbounded system.

\begin{figure}[htb]
\centerline{\includegraphics[width=6cm,height=6cm]{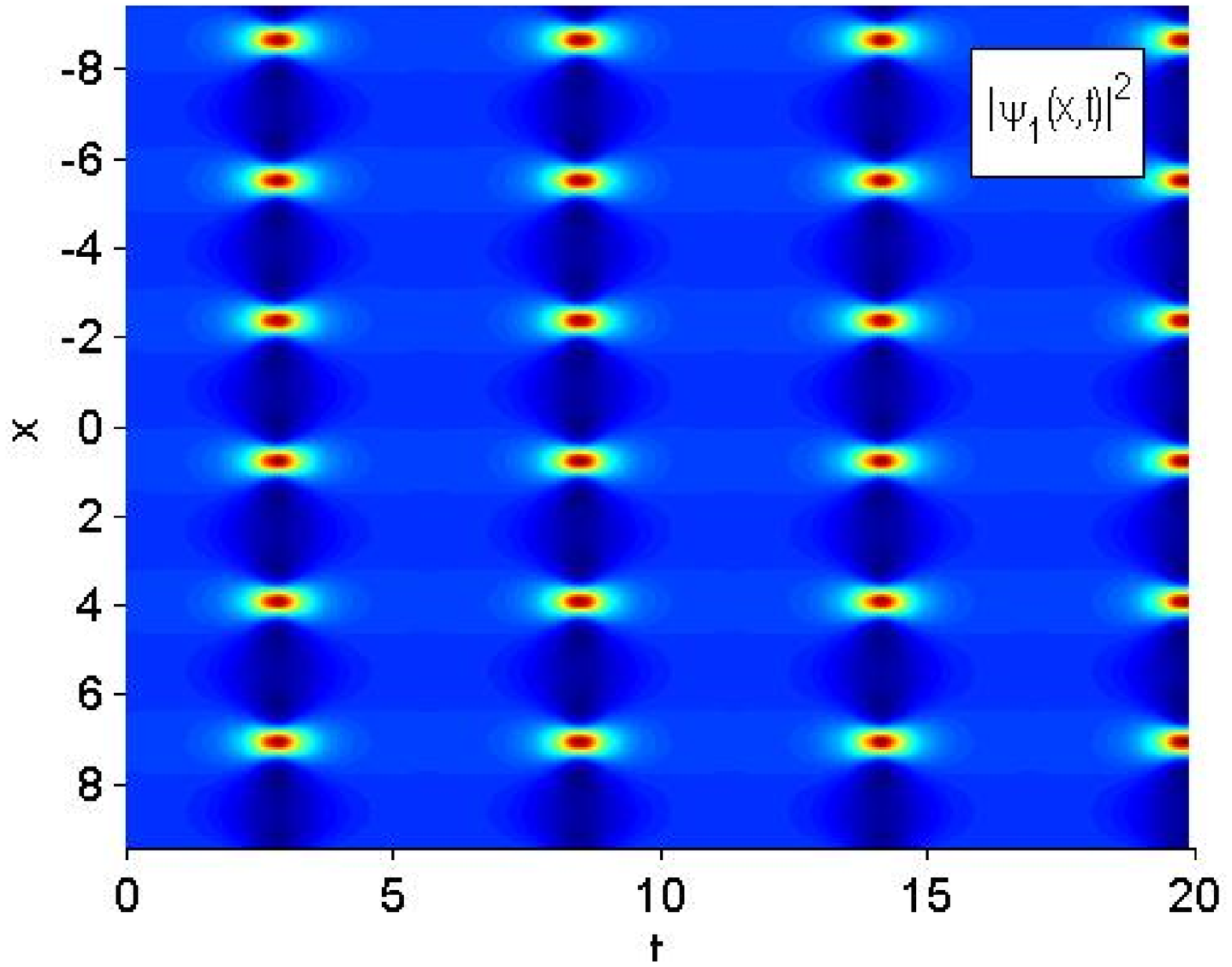}\quad
            \includegraphics[width=6cm,height=6cm]{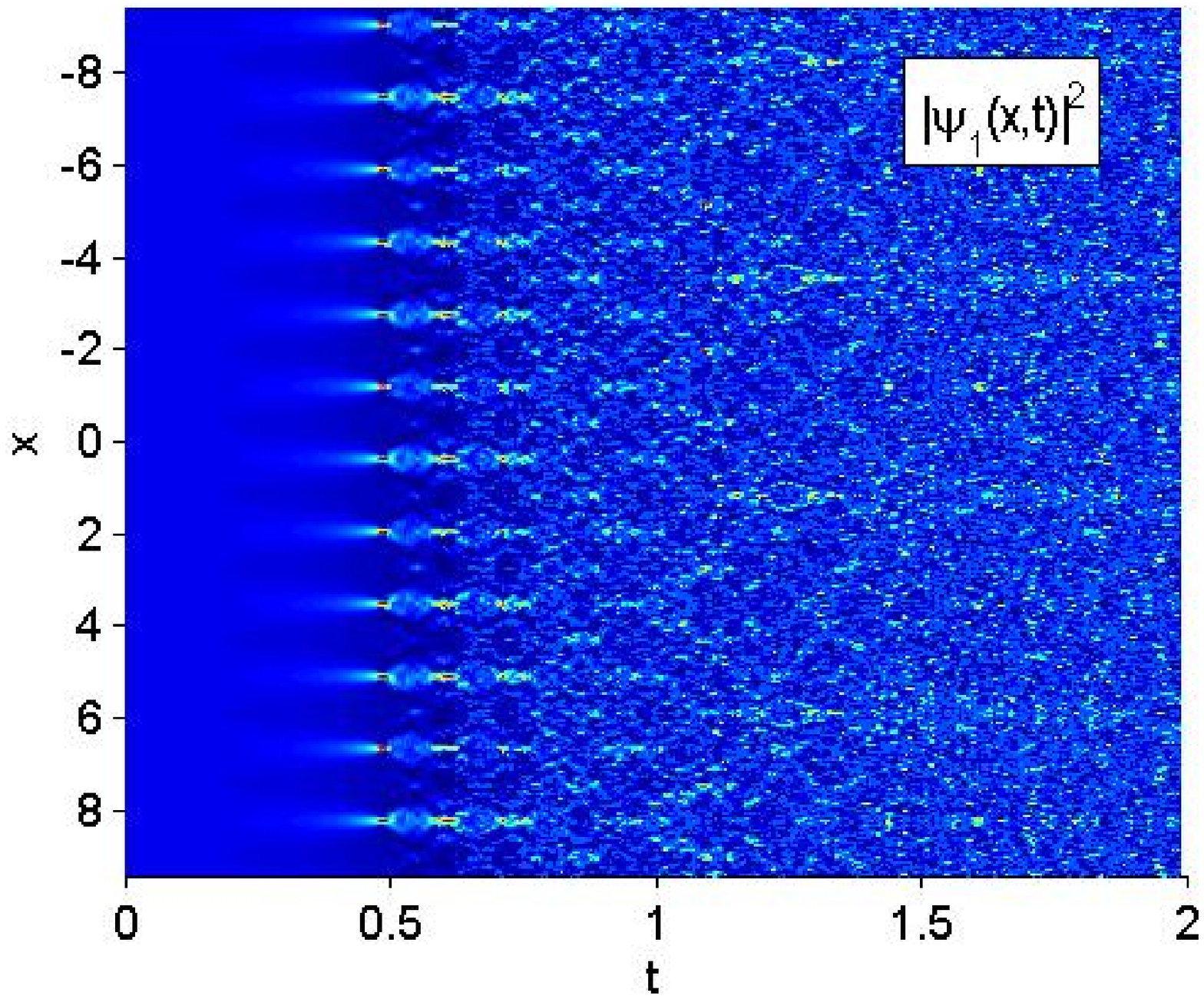}}
\caption{The results of numerical simulation of the governing Eqs.
(\ref{gpe}) with a weakly modulated plane wave initial conditions,
given by Eqs. (\ref{psi})-(\ref{pert}): $\psi_j(x) = a_j + \epsilon
\sin(q_j x)$; $a_1 = 1.0$; $a_2 = 0.999$; $\epsilon = u_j = v_j =
0.01$; $q_1 = q_2 = 2.0$; $k_j = 0$. Left panel: Only the cubic
nonlinearity is in action $\lambda=1$, $\beta=1$, $\gamma=0$. Right
panel: Both cubic and quintic nonlinearities are in action
$\lambda=1$, $\beta=1$, $\gamma=1$, $\alpha=1$, $q_1 = q_2 = 4.0$.}
\label{fig2}
\end{figure}

\begin{figure}[htb]
\centerline{\includegraphics[width=8cm,height=6cm]{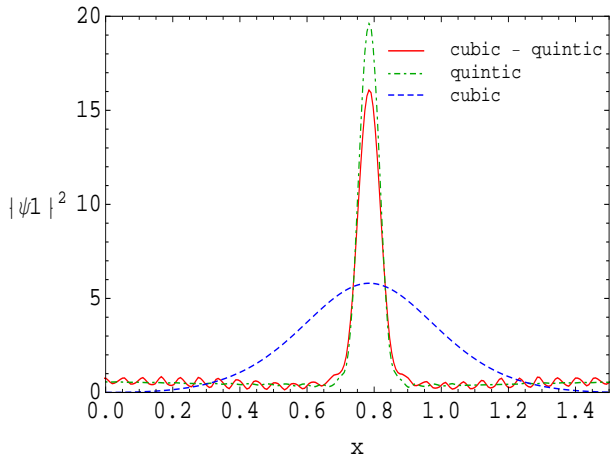}}
\caption{Solitons emerging from the MI when the instability is
caused by different types of nonlinearities (only one pulse of the
soliton train is shown for each type). } \label{fig3}
\end{figure}

In Fig. \ref{fig2} we present the results of numerical simulations
of Eqs. (\ref{gpe}) for different sets of parameters. When only the
cubic nonlinear terms in the equations are preserved ($\lambda = 1,
\gamma = 0$), periodic emergence of soliton trains in the system is
observed, as shown in the left panel. Experimental observation of
the associated Fermi-Pasta-Ulam recurrence phenomenon in propagation
of modulationally unstable waves in optical fibers with Kerr type
nonlinearity was reported in \cite{simaeys2001}. In contrast, when
both types of nonlinearity are taken into account ($\lambda=1$,
$\gamma=1$), the periodic emergence of soliton trains is compromised
(right panel). In this case the soliton trains clearly emerge only
two times, after that almost random distribution of pulse
intensities sets in. We have shown the evolution of one component
$\psi_1(x,t)$ of the system (\ref{gpe}), because the other component
$\psi_2(x,t)$ behaves similarly.
An important difference between the considered cases is that, the
quintic and combined cubic-quintic nonlinearities give rise to
solitons with bigger amplitude compared to those of cubic
nonlinearity, as shown in Fig.~\ref{fig3}. As expected, the higher
order nonlinearity produces strongly compressed solitons and leads
to cessation of the recurrence of soliton trains.

The linear stability analysis, described in the previous section, is
valid only for the initial stage of MI. When the amplitude of
perturbations becomes comparable with the background, the nonlinear
stage of MI sets in, for which adequate theory has not been
developed yet. However, some evidences are found for the universal
behavior of waves emerging from MI in all media \cite{biondini2016}.
In particular, when localized perturbation is imposed upon the
constant background, the evolving wave field distinctly divides into
two outer regions, where the wave amplitude is the same as in
unperturbed solution, and the central expanding oscillatory region.
As a signature of the influence of quintic term on MI, we look for
changes in this central oscillatory region of the integration
domain.

\begin{figure}[htb]
\centerline{\includegraphics[width=6cm,height=4cm]{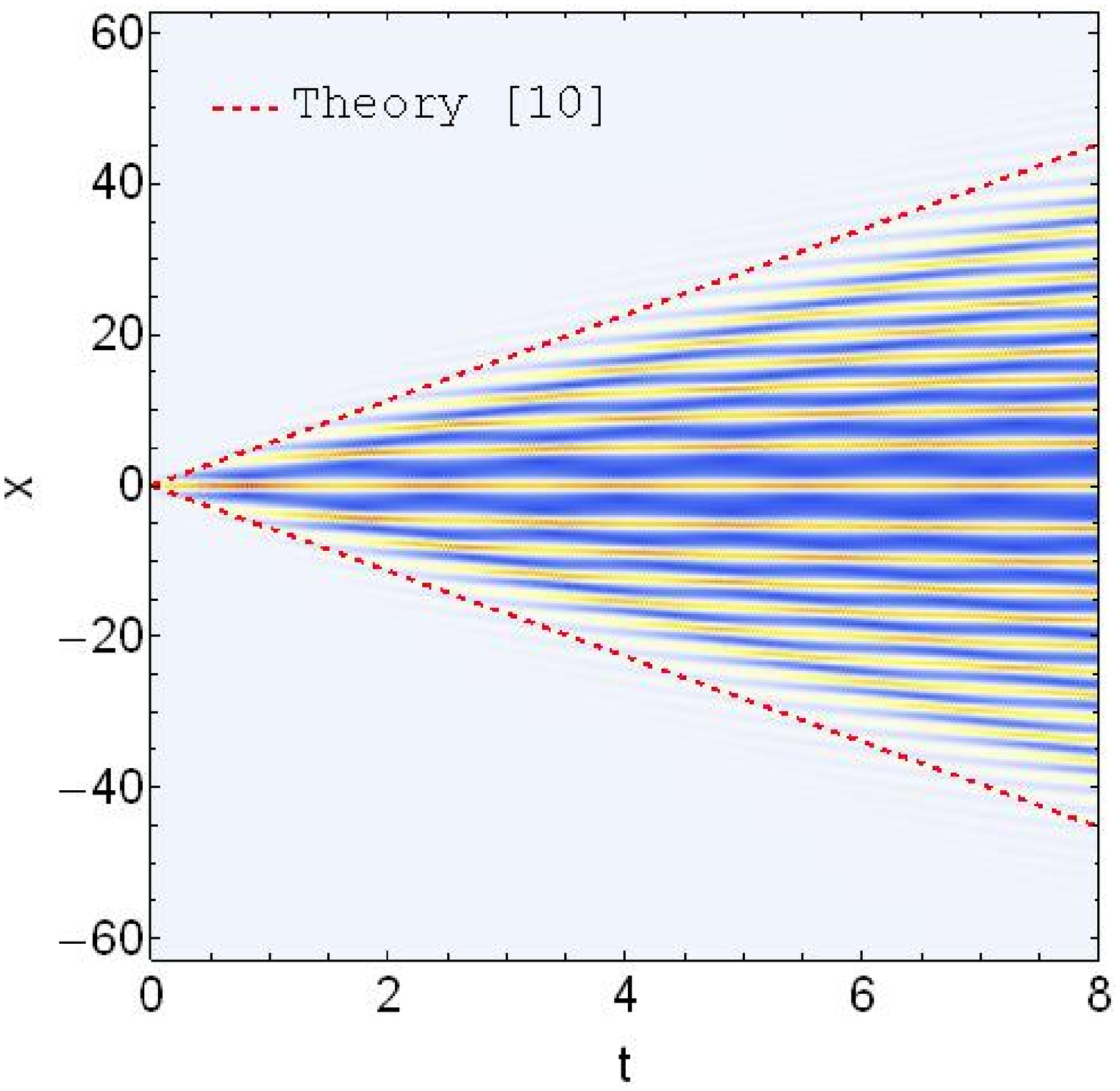}\quad
            \includegraphics[width=6cm,height=4cm]{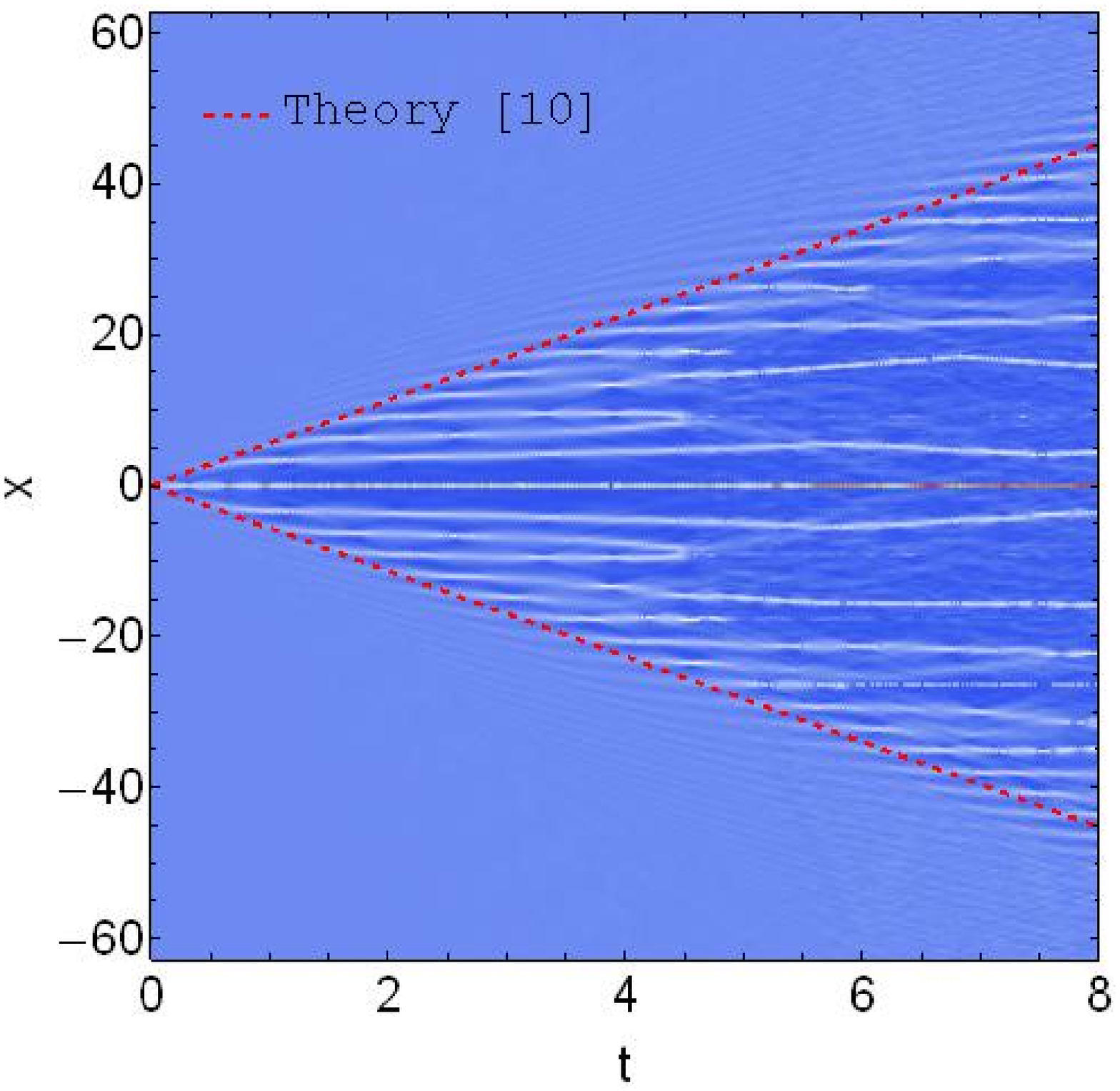}}
\vspace{0.25cm}
\centerline{\includegraphics[width=6cm,height=4cm]{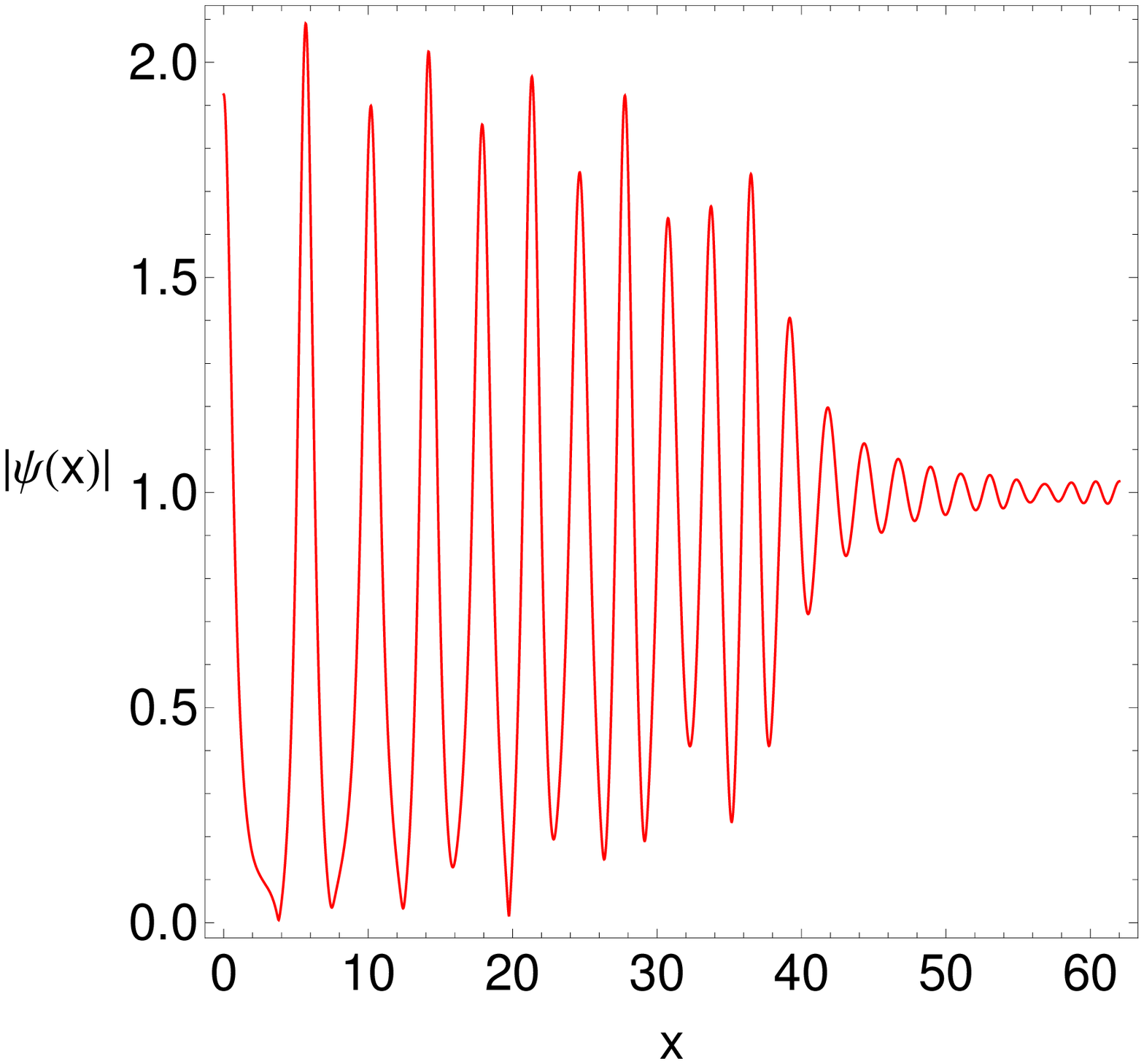}\quad
            \includegraphics[width=6cm,height=4cm]{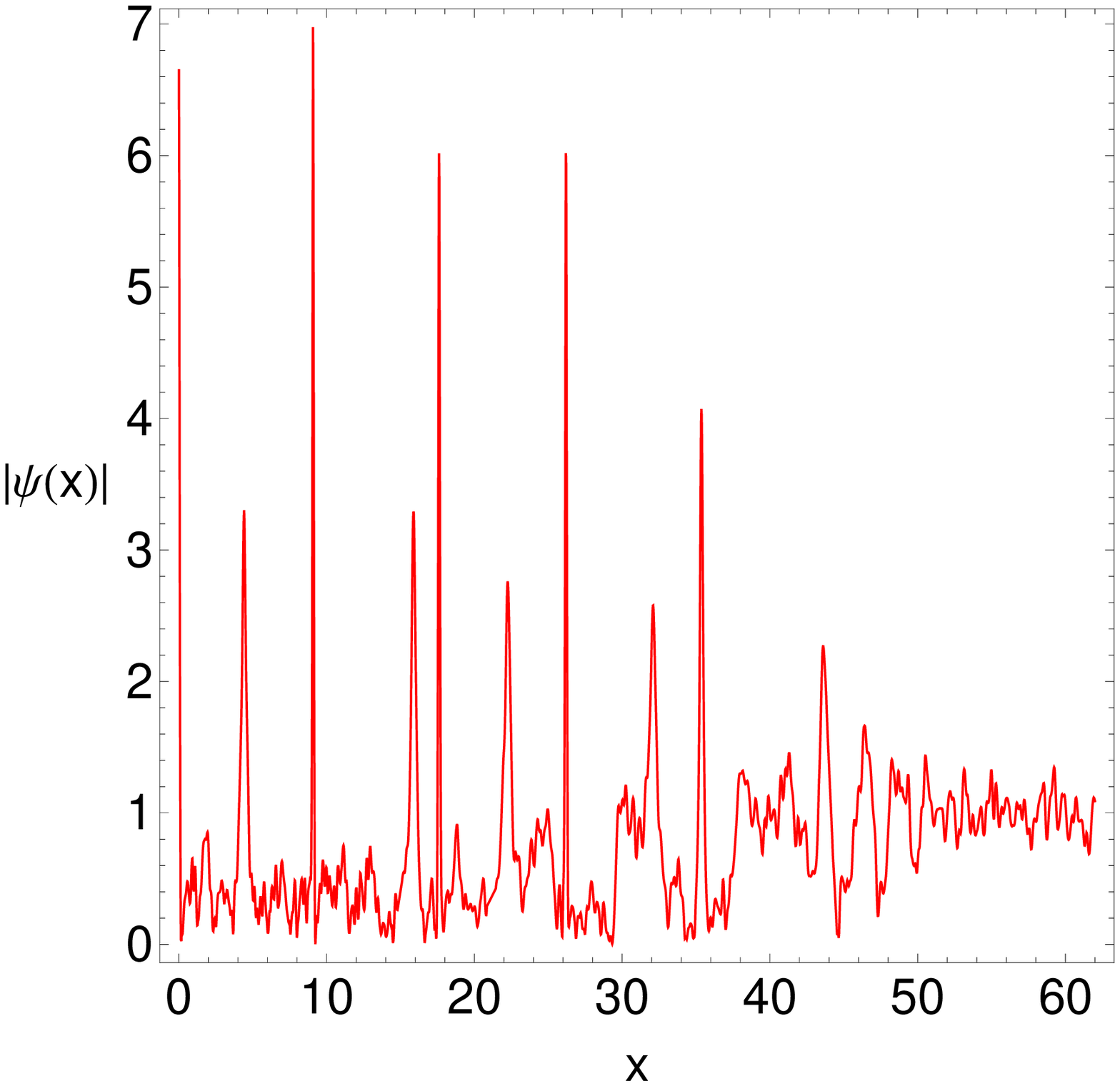}}
\caption{Top panels: Density plots of the wave field $|\psi(x,t)|$,
evolving from locally perturbed initial condition Eq.
(\ref{locally}) according to Eq. (\ref{gpe}). Left: For single NLSE
with cubic nonlinearity ($\lambda=2$, $\beta=0$, $\gamma=0$,
$\alpha=0$) the oscillatory central region expands as predicted by
theory \cite{biondini2016} $ \sim \pm 4\sqrt{2} \,t$ (red dashed
lines). Right: For single NLSE with cubic-quintic nonlinearity
($\lambda=2$, $\beta=0$, $\gamma=0.5$, $\alpha=0$) the oscillatory
region expands linearly with the same velocity, except emerging
solitons are more compressed. Bottom panels: The solution
$|\psi(x,t)|$ at $t=8$ for cubic (left) and cubic-quintic (right)
NLSE. } \label{fig4}
\end{figure}

For this purpose we impose a Gaussian localized perturbation on the
background as in Ref. \cite{biondini2018}
\begin{equation}\label{locally}
\psi(x,0)=1+i \exp(-x^2) \cos(\sqrt{2} x).
\end{equation}
Inserting this locally perturbed background solution as initial
condition into Eq. (\ref{gpe}), and propagating in time, we find
clear changes due to the quintic term. The results are shown in Fig.
\ref{fig4}.
The main difference in the nonlinear evolution of MI in cubic and
cubic-quintic NLSE is that, in the latter case the emerging solitons
are more compressed and evenly spaced (compare the bottom panels in
Fig. \ref{fig4}). However, the universal feature of nonlinear MI in
both media is preserved. Namely, the central oscillatory region of
the wave field expands linearly with time, and the outer quiescent
region is quite sharply separated from central oscillatory region of
MI.

\section{Coupled solitons emerging from MI}

In the absence of inter-component interactions ($\alpha = 0, \ \beta
= 0$) the system of Eqs. (\ref{gpe}) splits into two independent NLS
equations with cubic-quintic nonlinearity, which we present in a
more convenient form by setting $c_j \rightarrow 1/2$, and $\gamma
\rightarrow -\gamma$
\begin{equation}\label{cqnlse}
i\psi _{t}+\frac{1}{2}\psi _{xx} + \lambda |\psi |^{2}\psi -\gamma
|\psi |^{4}\psi  =0.  \label{gpe1}
\end{equation}
An essential property of Eq. (\ref{cqnlse}) is that it supports so
called flat-top solitons \cite{pushkarov1979,maimistov-book}
\begin{eqnarray}\label{flattop}
\psi(x) &=& \sqrt{\frac{3\lambda}{4\gamma}} \ \frac{{\rm
tanh}(\eta)}{\sqrt{1 + \mathrm{sech}(\eta) \mathrm{cosh}(x/a)}}, \\
\qquad \eta &=& \sqrt{\frac{2\gamma}{3}}, \quad a
=\frac{\eta}{\lambda \mathrm{tanh}(\eta)}, \nonumber
\end{eqnarray}
a pedestal-shaped localized states which can propagate preserving
their form and velocity, but collide with each-other inelastically
\cite{elyutin2009}.

Below we consider the MI in our system with flat-top soliton initial
conditions in both components. Such a setting with vanishing
boundary conditions is convenient for observing coupled solitons,
emerging from MI and evolving in the domain of integration for
extended period of time. Since the interaction between flat-top
solitons is a strong perturbation, substantial amount of radiation
of linear waves will take place during the process. These linear
waves have to be removed from the integration domain since they can
be reflected from the domain boundaries and hence interfere with
emerging structures in its central part. For this purpose we use the
technique of absorbing boundaries proposed in \cite{berg}. A similar
setting with vanishing boundary conditions was employed in
\cite{kivshar1991} to study the generation of symbiotic optical
solitons in coupled system of NLSE with Kerr type nonlinearity.

\begin{figure}[htb]
\centerline{\includegraphics[width=6cm,height=7.5cm,clip]{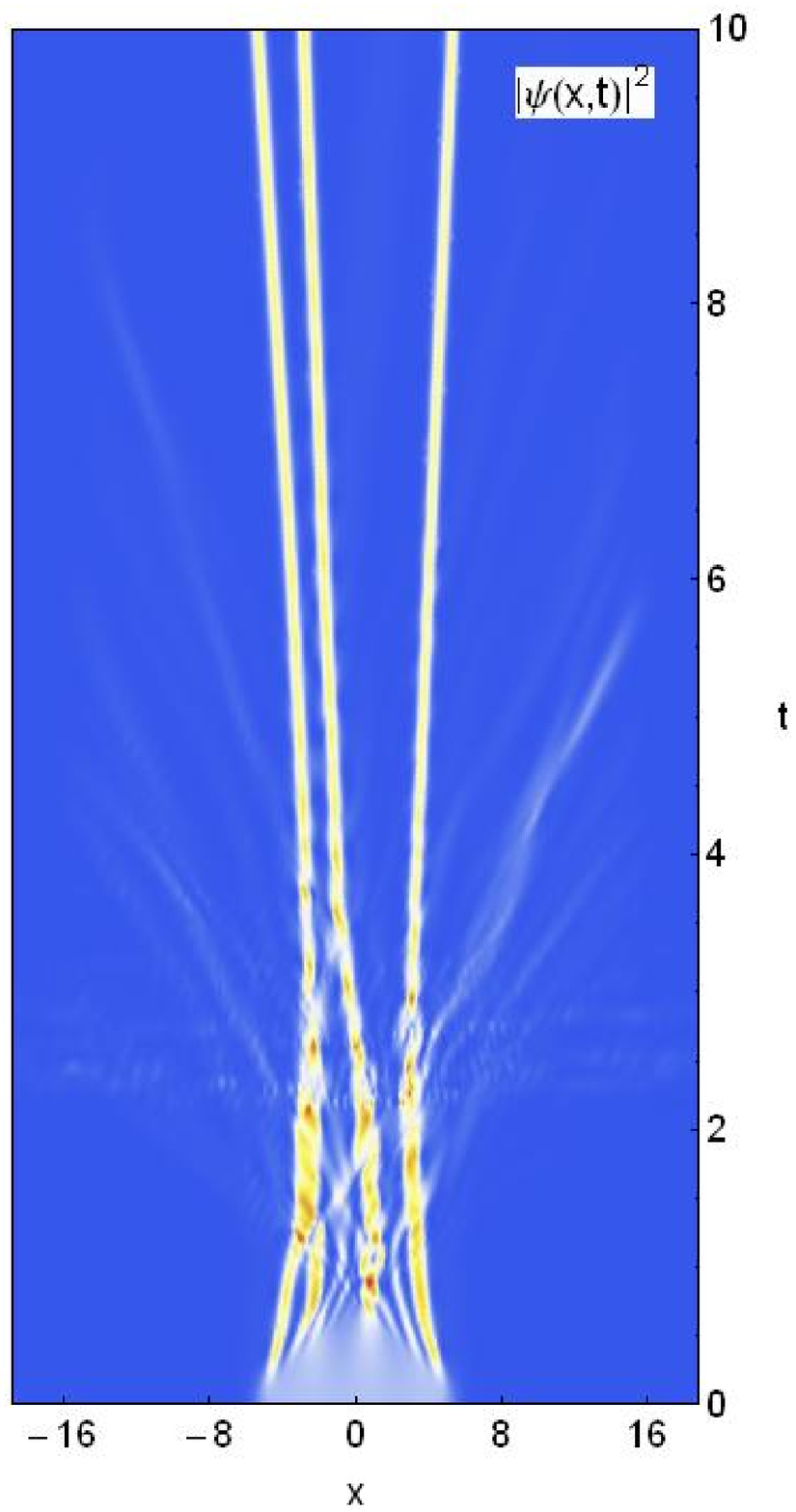}\quad
            \includegraphics[width=6cm,height=7.5cm,clip]{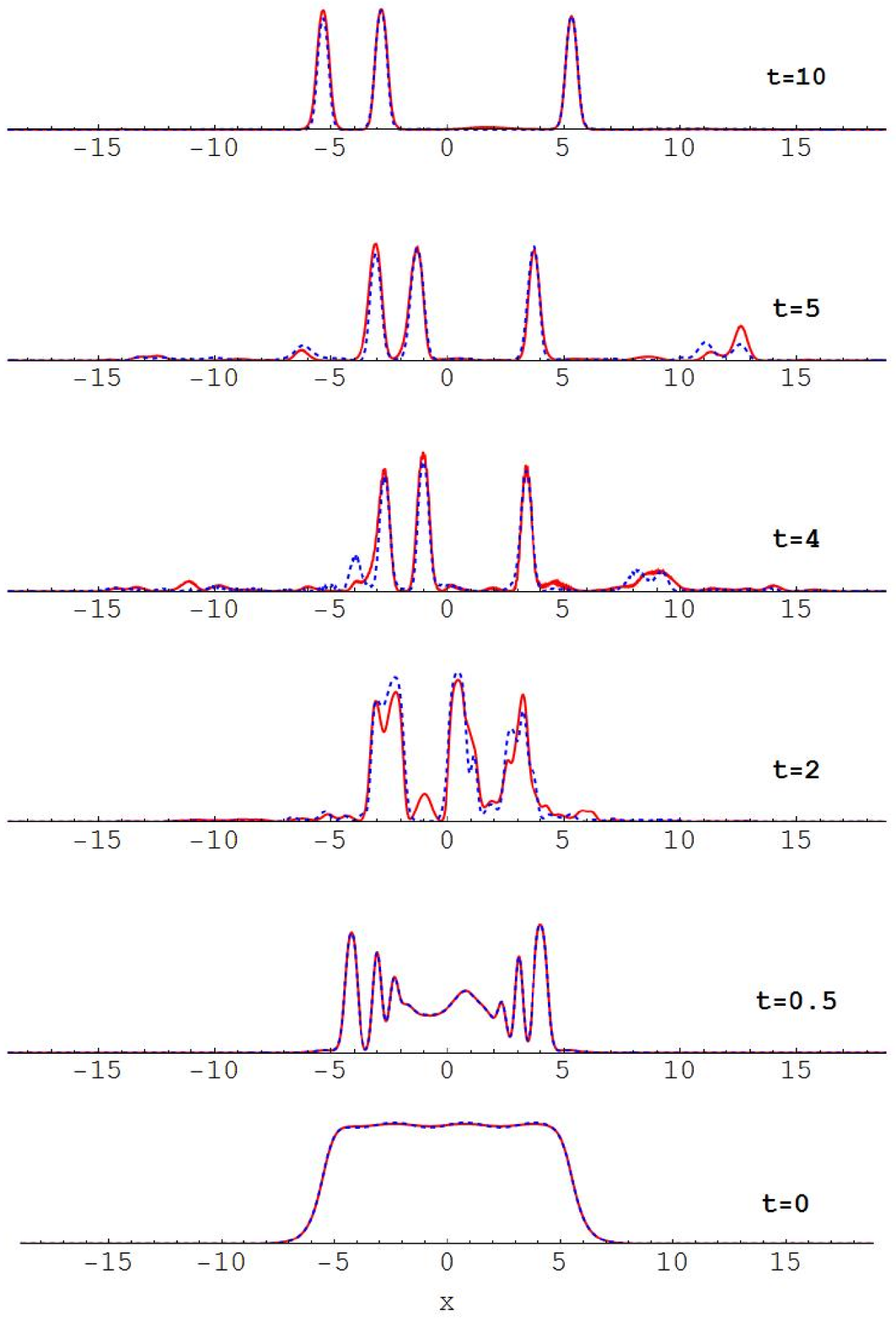}}
\caption{Formation of coupled solitons due to the phenomenon of MI
in NLS Eqs. (\ref{gpe}). The initial conditions are taken as weakly
modulated flat-top solitons Eq. (\ref{initcond}). Left panel:
Density plot for the time evolution of the first component
$|\psi_1(x,t)|^2$. The evolution of the second component is similar,
that is why not shown. Right panel: The initial state ($t=0$) and
several time sections for the evolution of the coupled system
(\ref{gpe}) under developing MI. Red solid line and blue dashed line
correspond to the first and second components, respectively.
Formation of three coupled solitons can be clearly observed. The
length of integration domain $L=24 \pi/q$ is selected to accommodate
an integer number of modulation periods. Parameter values: $c=0.5$,
$\lambda=100$, $\beta=1$, $\gamma=800$, $\alpha=-0.1$, $q=2$,
$\epsilon_1=0.01$, $\epsilon_2=0.02$. } \label{fig5}
\end{figure}

To start the numerical simulations we introduce the following
perturbed flat-top soliton initial conditions into Eqs. (\ref{gpe})
\begin{equation}\label{initcond}
\psi_j(x,0) = \psi(x) [1+\epsilon_j \sin(q x)], \quad j=1,2,
\nonumber
\end{equation}
with $\psi(x)$ given by Eq. (\ref{flattop}). These two flat-top
solitons are similar except for the weak perturbations with
different amplitudes ($\epsilon_j$) imposed on them. The evolution
of two interacting pulses~(\ref{flattop}), governed by Eqs.
(\ref{gpe}) is presented in Fig. \ref{fig5}. In order to have
extended pedestal shaped pulses with sufficient background
intensity, we use suitable values for the parameters $\lambda$ and
$\gamma$ of the flat-top soliton. Estimation of these parameters for
BEC of $^{87}$Rb can be found in \cite{baizakov2011} As can be seen
from this figure, MI fully develops around $t \sim 0.5$, after that
emergence of coupled solitons takes place in the interval $t \in [2,
5]$. Three coupled solitons fully developed and freely propagate
when $ t > 5$. It should be noted, that formation of coupled
solitons is possible when the norm of each soliton is big enough so
that mutual attraction between them overcomes the quintic
self-repulsion ($\lambda > 0$, $\beta > 0$, $\gamma < 0$, $\alpha <
0$). For this reason smaller pulse pairs to the left and right of
the three central pulses disperse and leave the integration domain
by $ t \sim 10$.

\section{Conclusions}

We have studied the phenomenon of MI in a two-component continuous
system featuring cubic-quintic nonlinearity both in linear and
nonlinear stages. For the linear stage analytic expression for the
growth rate of MI has been derived and compared with the results of
numerical simulations of the governing NLSE system. The model allows
to identify the contribution of each type of nonlinearity on the
overall growth rate of MI. It is found that the quintic nonlinearity
significantly enhances the development of MI in these media and
suppresses recurrence property of emerging soliton trains. To
investigate MI in the nonlinear stage we use the evidence of a
universal behavior, discovered in Ref. \cite{biondini2016}. It
appears, that the quintic nonlinearity leads to emergence of more
compressed and evenly spaced solitons, while the expansion of the
central oscillatory region obeys the same linear time dependence. In
addition generation of coupled solitons of the bright-bright type
resulted from MI of flat-top solitons has been investigated via
numerical simulations. Obtained results can be useful in studies of
binary mixtures of BEC with high density, where the three-body
atomic interactions play a significant role, and of light
propagation in optical materials with a large fifth-order
nonlinearity.

\section*{Acknowledgements}

This work has been supported by the KFUPM research group projects
RG1503-1 and RG1503-2. BBB thanks the Physics Department at KFUPM
for their hospitality during his visit.

\end{document}